# A Generic Trauma Severity Computer Method Applied to Pedestrian Collisions

Christophe Bastien[1], CIive NeaI-Sturgess[2], Huw Davies[1]

1. Coventry University, Institute for Future Transports and Cities, Coventry, UK
2. University of Birmingham, Department of Mechanical Engineering, Birmingham, UK

**Abstract:**
In the real world, the severity of traumatic injuries are measured using the Abbreviated Injury Scale (AIS). However the AIS scale cannot currently be computed by using finite element human computer models, which calculate a maximum principal strains (MPS). Further, MPS only establishes a threshold above which a serious or fatal injury occurs. In order to overcome these limitations, a unique Organ Trauma Model (OTM) able to calculate the threat to life of any organ injury is proposed. The focus, in this case is on real world pedestrian brain injuries. The OTM uses a power method, named Peak Virtual Power (PVP), and defines brain white and grey matters trauma responses as a function of impact location and impact speed extracted from the pedestrian collision kinematics. This research has included ageing in the injury severity computation by including soft tissue material degradation, as well as brain volume changes. Further, to account for the limitations of the Lagrangian formulation of the brain model in representing haemorrhage, an approach to include the effects of subdural hematoma is proposed and included as part of the OTM predictions in this study. The OTM model was tested against three real-life pedestrian accidents and has proven to reasonably predict the Post Mortem (PM) outcome. Its AIS predictions are closer to the real world injury severity than standard MPS methods currently recommended. This study suggests that the OTM has the potential to improve forensic predictions as well as contribute to the improvement in vehicle safety design through the ability to measure injury severity. This study concludes that future advances in trauma computing would require the development of a brain model which could predict haemorrhaging.

## *1. Introduction*

Automotive manufacturers design vehicles to meet legislative and consumer test protocols using anthropometric crash test devices (ATD) with the purpose of creating safer vehicles for both occupants and pedestrians. In spite of all their efforts, the number of fatalities keeps on increasing worldwide year by year [1], reaching 1.35 million in 2018. There are many parameters which can be attributed to this increase of death toll such changes to age, gender, speeding, infrastructure etc..., however, the steady rise in numbers begs the question whether the design tools currently used in the design process namely crash test dummies, are adequate to reverse this trend. ATD record displacements, accelerations and forces. During the vehicle design process, the ATDs output information is cross-correlated to a probability of threat to life, based on injury severity, defined by medical professionals who have suggested a trauma injury scale or the Abbreviated Injury Scale (AIS) [2]. The AIS is internationally accepted and is the primary tool to conclude injury severity and is anatomically based. It is a, consensus derived, global severity scoring system that classifies each injury by body region according to its relative importance (threat to life) on a 6-point ordinal scale and provides a standardised terminology to describe injuries and ranks injuries by severity. The measurements from crash test dummies can only be used to speculate on the probability of death and have no internal organs, consequently they are not useful in predicting soft tissue injuries in a deterministic manner. Human computer models, like the THUMS [2], have modelled the soft organ tissues (heart, kidneys, liver, spleen, liver, grey and white matter) and can output soft tissue Maximum Principal Strains (MPS), which unfortunately only have a bearing with AIS4 [4][5][6]. MPS are standard outputs suggested by Human Computer models, which is a major limitation in injury severity computation prediction. This paper proposes a new Organ Trauma Model (OTM) to compute soft tissue trauma. This OTM model will be compared with the MPS method in the case of pedestrian collisions for which collision details as well as PM information have been provided by the UK Police Force (UKPF).

---

[1] Corresponding author: aa3425@coventry.ac.uk

## 2. Derivation of an Organ Trauma Model (OTM)

A mathematical derivation was performed to link threat to life the results of a Finite Element Analysis (FEA) computation focused on a vehicle to pedestrian collisions. One of the innovations and challenges of this research was the coding of trauma and poly-trauma in a computer simulation. As there is no direct link between Maximum Principal Strains (MPS) and injury severity, it was proposed to use the Peak Virtual Power (PVP) theory applied to soft organ tissues to compute their injury severity [4][5][6].

2.1 Theoretical Derivation of Trauma

When a human is impacted on the thorax, the rib cage deforms under load, causing the organs to experience strain due to the force which is applied to them. The event can be very sudden in the case of a pedestrian-to-vehicle collision (Figure 1), the primary impact lasting in the order of 0.1s to 0.2s depending on the vehicle impact speed.

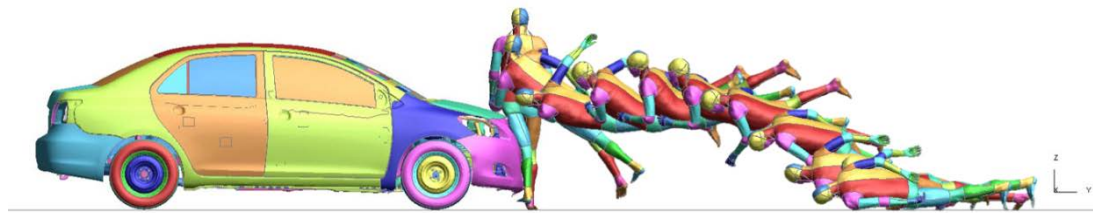

*Figure 1: Typical pedestrian kinematics during collision with braking*

Injury severity can be computed from a concept called Peak Virtual Power (PVP). Peak Virtual Power is based on the general principle of the 2nd law of thermodynamics, stating that entropy (state of disorder) increases after each mechanical process. When a collision takes place, the entropy (represented by PVP) always increases, never to return. A typical pattern of this behaviour is illustrated in Figure 2, where organ power goes up and down, while PVP keeps always to the maximum value at all times.

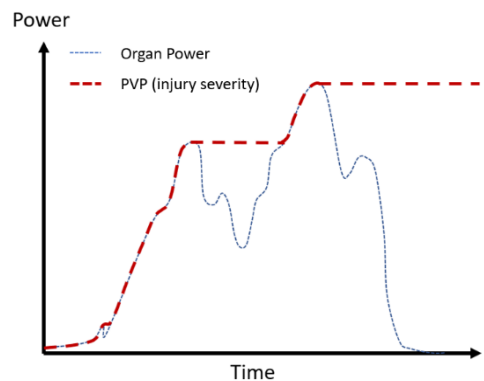

*Figure 2: Power in an organ goes up and down, while trauma (represented by PVP) keeps on increasing [9]*

PVP in a finite volume of the body (at organ level for example) is calculated by multiplying the localised Von Mises stress in that volume ($\sigma_{VM}$), by its speed of deformation (or Von Mises strain rate ($\dot{\varepsilon}_{VM}$)). As the load varies during the impact, organ power will vary while PVP will always take the maximum value (Figure 2).

It is demonstrated that the resultant injury severity is a consequence of this increase of entropy and is proportional to the PVP generated by this collision (Equation 1). If PVP increased, then the trauma injury increases [4][5][6].

$$PVP \propto max(\sigma \cdot \dot{\varepsilon}) \propto AIS$$

*Equation 1: Generic relationship between Peak Virtual Power and threat to life.*

The injury severity is coded via an Abbreviated Injury Scale (AIS), which has been medically derived and listed in Table 1.

| AIS Level | Injury | Risk of death % |
| --- | --- | --- |
| 1 | Minor | 0.0 |
| 2 | Moderate | 0.1 -0.4 |
| 3 | Serious | 0.8 – 2.1 |
| 4 | Severe | 7.9 – 10.6 |
| 5 | Critical | 53.1 – 58.4 |
| 6 | Un-survivable | 100 |

*Table 1: Abbreviate Injury Scale linking AIS level and risk to life [10]*

When using human computer models in accident reconstruction, it is possible to relate the threat to life to human organ tissue deformations observed during real human organ tests. If the maximum principal strain threshold (MPS) is exceeded then severe injuries will occur (usually an AIS 4 outcome). The list of cut-off injury values, used in this study, are listed in Table 2. If some zones in the white matter stretches by 21% (computed), then Diffuse Axon Injuries (DAI) will occur (AIS 4), as seen in Table 1.

| **Body Part** | **Load** | **Threshold** | **AIS level** |
| --- | --- | --- | --- |
| **Brain contusion** | Maximum principal strain | 26% [11] | 3 |
| **Diffuse Axonal Injury (DAI)** | Maximum principal strain | 21% [12] | 4 |

*Table 2: Injury trauma values used in THUMS [15]*

By considering further scientific literature [17][18], it was observed that the threat to life increases by a cubic relationship when AIS is increased (Figure 3).

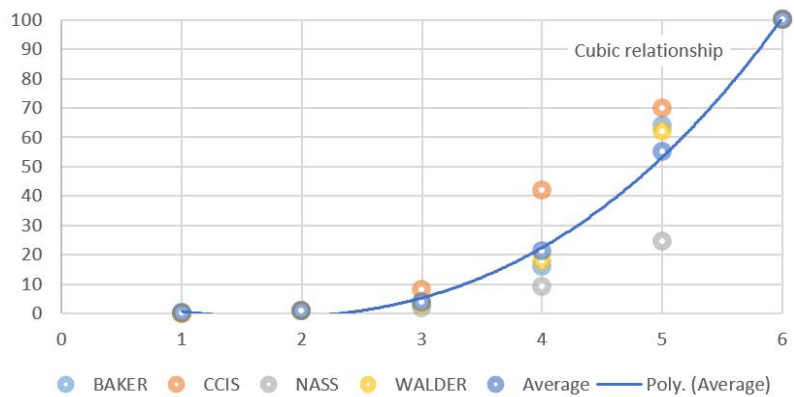

*Figure 3: Relationship between threat to Life and AIS [17][18]*

This is a very important observation, as if the PVP necessary to cause a severe injury is known (AIS 4) then it is possible to extract how much PVP the organ can withstand to reach AIS 1, 2, 3 and 5. The PVP values can be scaled from AIS 4 by the ratios $1^3/4^3$, $2^3/4^3$, $3^3/4^3$ and $5^3/4^3$ respectively to create the full map of trauma injuries for that organ, creating an "Organ Trauma Model" (OTM).

As an illustration, any OTM, will be therefore represented by a graph containing the relationship between PVP, impact velocity and AIS, as illustrated in Figure 4. It has been possible to include error corridors (upper and lower) for each AIS value by considering the spread of data from Figure 3.

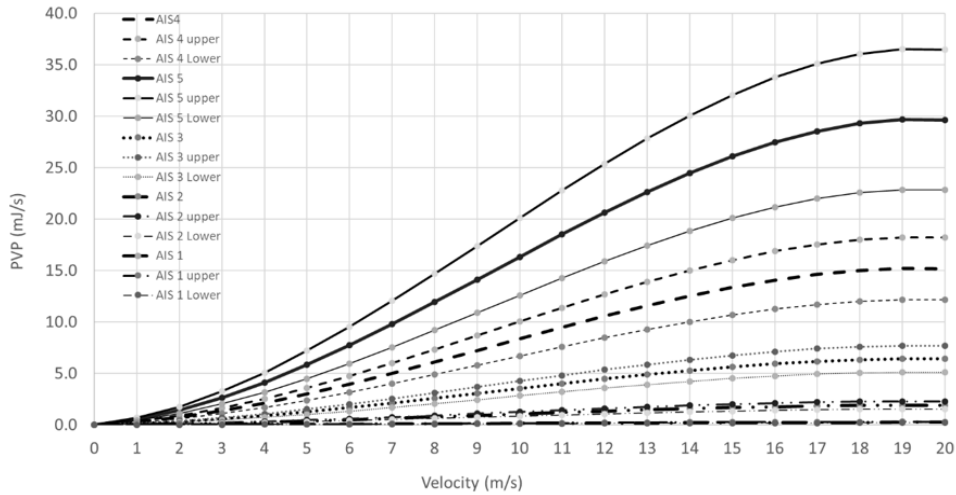

*Figure 4: Organ Trauma Model (OTM) for a head impact of the forehead against a rigid impactor [10]*

As an example, looking at Figure 4, the following arbitrary scenarios can be concluded (Table 3):

| Impact speed (m/s) | PVP (mJ/s or mW) | AIS extracted from Figure 4 |
| --- | --- | --- |
| 9 | 7.5 | 4 |
| 14 | 25 | 5 |
| 19 | 10 | 3 or 4 (depending on how close the PVP value is from the upper AIS 3 and lower AIS 4 corridors |

*Table 3: Hypothetical scenarios extracted from Figure 4*

In order to understand the key parameters influencing PVP, and therefore the trauma severity, it is necessary to 'rework' Equation 1 into a more arithmetical form.

2.2 Algebraic formulation of Trauma Severity

By equating the organ kinetic energy and its deformation energy during the impact, it can be shown that AIS depends on the geometry of the organ at the time of impact, its material properties, the stiffness of the impacted surface and the velocity cubed (Equation 2). The whole derivation, validation and justification of Equation 2, is given in Appendix 1 and Appendix 2.

$$AIS \propto PVP = \frac{A_p}{2V_p} \sqrt{\left(\frac{\rho_p E_p \rho_c E_c m_p}{\rho_p E_p m_c + \rho_c E_c m_p}\right)} v_{t_0}^{\ 3}$$

*Equation 2: Generic Algebraic derivation of PVP*

<u>Where:</u>

- '$A_p$' represents the contact of the Area of the organ which is impacting the vehicle. This Area will change according to the kinematics of the pedestrian while wrapping around the vehicle profile
- '$V_p$' is the volume of the organ (constant)
- '$\rho_p$' is the density of the organ
- '$\rho_c$' is the density of the contact surface
- '$m_p$' is the organ mass

- '$m_c$' is the vehicle mass
- '$E_p$' represents the Modulus of Elasticity of the organ (Young's / Bulk Modulus)
- '$E_c$' represents the stiffness of the vehicle
- '$v_{t0}$' is the organ impact speed, which is not necessary the vehicle impact speed. For an upright vehicle, i.e. bus, the organ impact speed is the bus impact speed, while in a low fronted vehicle, the speed of every part of the body do not impact the vehicle at the vehicle impact speed; these can be lower or higher. Such velocities can be computed during the accident reconstruction phase.

The outcomes of Equation 2 are sensible, as:

- The higher the impact speed 'v', the higher the injury.
- The stiffer the contact stiffness, the higher the injury.
- The heavier the contacted object, the higher the injury.

An important fact is that, because the phenomenon is related to impact mechanics, the stress wave travels through tissues differently according to which part of the human is impacted. Consequently, PVP, and therefore AIS, is impact direction dependant. As an example if a head is dropped on a rigid surface, the trauma will be different depending on the contact point (forehead, temple or occipital). In such a scenario, head injuries will be lower on the forehead than the temple and occipital for a given impact speed.

Another important point to notice that, in Equation 2, $V_0$ (organ volume) is constant. The method used to reconstruct the accidents is using finite elements. As a general principle, finite elements discretise the problem in small elements which are connected to each other, so the sum of these elements represent the whole problem. By cutting the problem in small parts, it is possible to investigate what can happen locally: this method is used to analyse complex shapes which differ greatly from say simple beams or plates which have been solved by engineers. Usually, organs which have a three dimensional aspect are represented by connected cubes (hexahedrons) or triangular based pyramids (tetrahedrons). This is the case with the computer model used in this study (THUMS 4.01). During the impact, these elements deform, stretch and change shape, however their volume remains constant. It is called a "Lagrangian" representation of the problem. The consequence, is that, should bleeding occur in the real-world accident, i.e. loss of volume due to the blood escaping the organ, then the finite elements will not be able to capture this. This is an inherent limitation which became apparent upon the derivation of Equation 2. Looking at Equation 2, should bleeding occur, then $V_0$ will reduce. As a consequence PVP, and consequently AIS, will increase, which is as expected. On the other hand, should bleeding not been observed, then Equation 2 should provide the correct answer. In order to investigate bleeding, it is proposed to include the effects of Subdural Hematoma (SDH), which has been defined for an MPS value of 25.5% [7]. The problem then is to assert the AIS outcome from bleeding, as a small bleed could add '1' AIS level to the current trauma severity computed or '2' if the bleeding is judged to be important by the pathologist [8]. In some cases, the quantity of blood loss could be subjective, hence for the purpose of being consistent and conservative, all instances of blood loss for the purpose of this study have a '+1' AIS increment on the base AIS computed. This methodology, used on falls, was previous published [23], however it was never used in pedestrian collisions.

2.3 Brain Volume Adjustment for Ageing

The human brain is the central organ of the human nervous system and consists of the white matter and the grey matter, the brain stem and the cerebellum [19]. Previous work has generated a regression relationship linking brain volume and age [20], which is illustrated in Equation 3.

$$V_{age} = -0.0037 * age + 1.808$$

*Equation 3: Relationship between age and volume loss*

In the model used in this study, the brain white and grey matter were scaled about the brain centre of gravity to adjust for ageing.

2.4 Ageing Coding

It has been evidenced that as people age, the frailer they become [19]. It can be therefore assumed that material properties are decreasing as a function of ageing. Equation 2 can be modified to highlight which terms are age dependant (Equation 4).

$$AIS \propto PVP = \left[\frac{A_p}{2V_p}\sqrt{\left(\frac{\rho_p E_p \rho_c E_c m_p}{\rho_p E_p m_c + \rho_c E_c m_p}\right)}\right]_{AGE} \cdot v_{t_0}^{3}$$

*Equation 4: Relationship between Trauma and ageing*

The human head consists of a fleshy outer portion surrounding the bony skull, within which sits the brain. A previous study [19] has highlighted significant cortical thinning in the outer and inner tables of the frontal, occipital, and parietal bones of females, predicting a loss between 36% and 60% of the original bone thickness from age 20 to 100 years. Cortical thickness changes in the males were found to be insignificant. However, it is the decline in bone quantity and quality that increases fracture risk in a progressive manner [22]. It was found that loss of bone thickness, material elasticity and density were key outcomes of ageing (Figure 5). It can be observed that the mechanical properties of a male have indeed reduced by 20% when the pedestrian is 80 years old, compared to a 20 year old pedestrian.

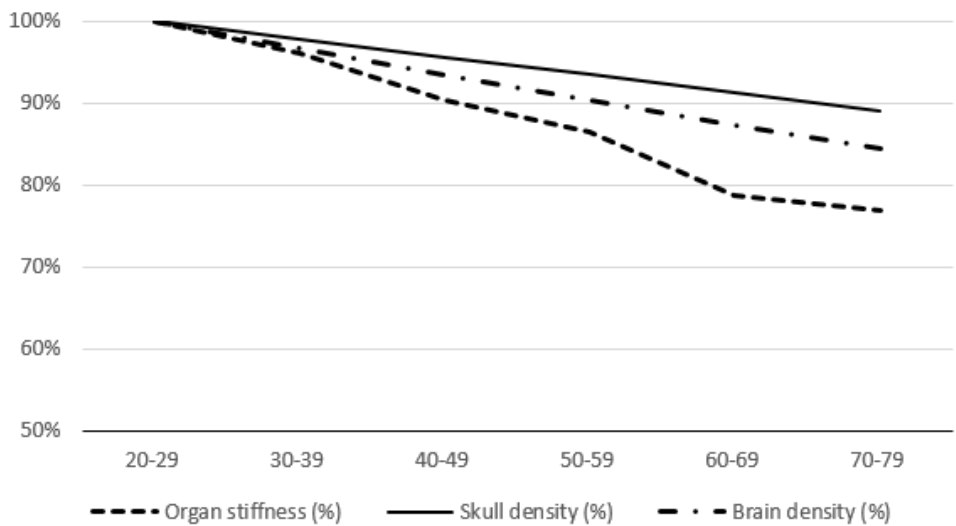

*Figure 5: Male Bone and Organ Performance as function of ageing [19][21][22]*

2.5 A Further Approach To The Model

This study was initially based on THUMS 4.01. When THUMS 4.02 was released, its main improvement was the white matter and grey matter material properties, which were publicised to be more representative [15], as well as a finer mesh. As the mesh density of THUMS 4.01 (3mm) was already suitable to observe an injury as part of a PM, it was decided to transfer the material properties of THUMS 4.02 into THUMS 4.01 and leave its mesh unchanged. Leaving the mesh unchanged has also the advantage to keep the runtime unchanged.

This was a choice of the authors, which has no influence on the methodology provided in the next session. The method provided is valid for any human computer model, like Global Human Body Model Consortium model (GHBMC), or any other model.

2.6 Testing the head model response.

For each accident that will be studied, the modulus of elasticity, density for all the soft tissue organs and bones, as well as the volume (for the brain only) will be adjusted to reflect the age of the pedestrian at the time of collision.

In order to illustrate the outcomes of Equation 4, when a human head computer model was impacted by an impactor on the forehead, it was noticed that it took less power for an older person to experience a head injury (Figure 6), at a set impact speed, which is consistent with what is observed in real life.

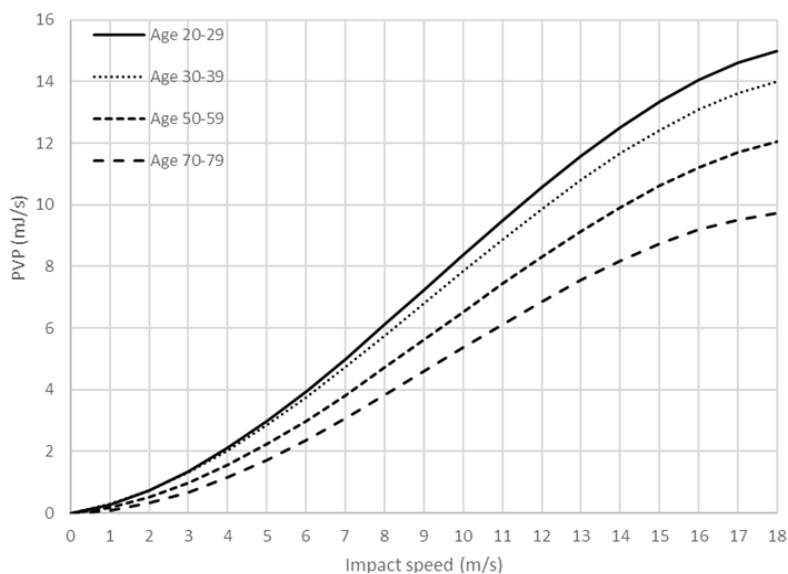

*Figure 6: Change of trauma response as a function of age. Each curve depicts a severe injury (AIS 4) [10].*

The mathematical model is consistent with the trends observed in real-life. It will be now tested against three real world pedestrian accident scenarios.

## 3. Methodology

The aim of the methodology section is to set out a framework to test the OTM model derived in Section 2. This methodology will consist of three steps which are illustrated in Figure 7:

- The first step is the accident reconstruction phase, whereby three accidents provided by the UK Police Force (UKPF) will be modelled. This accident reconstruction will recapture the collision event, by creating vehicles from their blueprints. These vehicles will be split as per their EuroNCAP pedestrian stiffness zoning [16] which will individually be represented by stiffness characteristics matching their real world test performance [16] . The pedestrians will be aged by scaling their mechanical properties, as per Figure 5, and sized and massed to their anthropometry. Once the accidents are computed, the full kinematics are extracted and compared to the damage observed (denting or smudge) on the vehicles to ensure that that

- the reconstruction is correct. Following this verification, the PVP values per organ for each collision as well as well as each organ velocity just before the impact are extracted.
- The second step is the trauma calibration phase. Considering Equation 4, the OTM model is specific for one vehicle only, for one point of impact only. Consequently, the OTM model has to be based on the same contact point. The full pedestrian kinematics from step 1 will be 'rewound' and the pedestrian is be positioned few millimetres from the bonnet (typically 3.0 mm). The direction of the velocity of impact vector from Step 1 are used then to impact the pedestrian at various velocities to construct an OTM model as Figure 4. It was checked that the head impact location was reasonably constant and it was observed that the variation in head impact location only varied by 4mm, which is negligible when compared to the size of the impact area. Consequently, the approach undertaken is compatible with keeping the impact location constant.

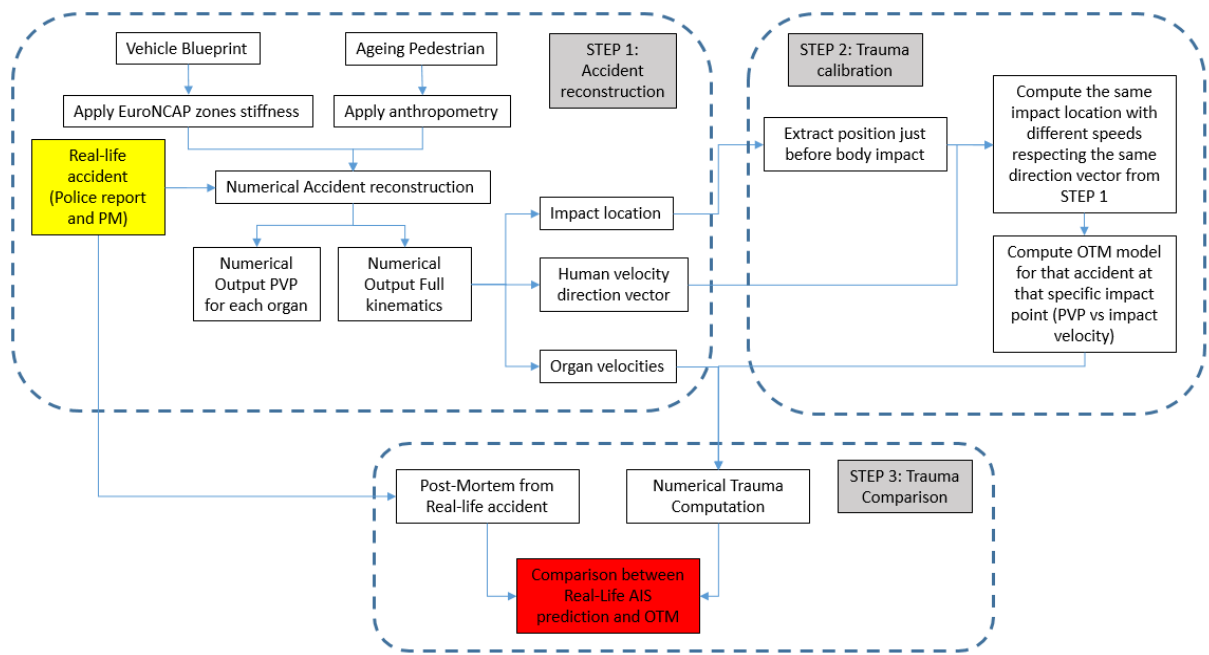

*Figure 7: Methodology to test the Mathematical OTM Model*

The velocity of interest is the impact velocity perpendicular to the windscreen, which is the main contributor to the blunt trauma impact. As such all velocities extracted in global coordinates (aligned with the vehicle) have been converted into windscreen coordinates (Table 4).

| Vehicle | Windscreen angle (°) |
|---|---|
| Renault Clio | 28.4 |
| Toyota Corolla | 23.0 |
| Mercedes Benz | 28.9 |

*Table 4: Windscreen angle relative to the horizontal plane at the head impact point*

Finally, step three will use the PVP and true brain impact velocity (perpendicular to the windscreen) responses from the first step and the OTM model built in the second step to propose a predicted AIS value. This AIS value will be compared to the value obtained in the real-life scenario from the post mortem. It is proposed that the OTM model is valid if both values have the same or similar AIS ordinal values.

## *4. Results*

4.1 Accident Reconstruction

Three accidents provided by the UKPF force were reconstructed. The details of each accidents are listed in Table 5 and the pedestrian damage and kinematics in Table 6.

| Case Id | Vehicle | Pedestrian Mass (kg) | Pedestrian height (m) | Age (year) | Impact direction | Vehicle Impact Speed (m/s) |
|---|---|---|---|---|---|---|
| 1 | Toyota Corolla | 58.6 | 1.65 | 34 | Right side impact (right leg forward) | 11.2 |
| 2 | Renault Clio | 79.2 | 1.73 | 79 | Side (left leg forward) | 12.5 |
| 3 | Benz B180 | 56.4 | 1.65 | 25 | from driver's near to far side | 12.5 |

*Table 5: UKPF Cases studied*

| Case id | Pedestrian Kinematics | Vehicle Damage |
|---|---|---|
| 1 | 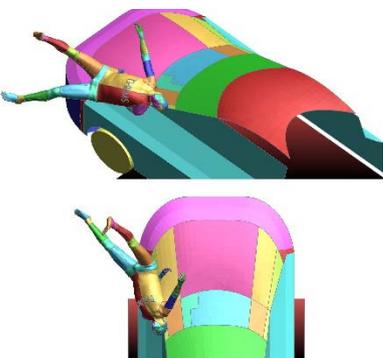 | 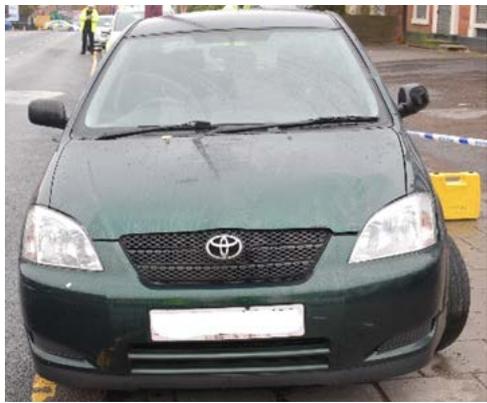 |
| 2 | 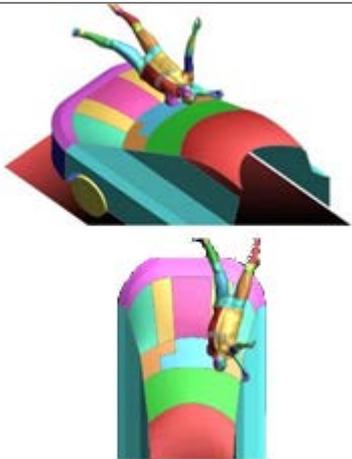 | 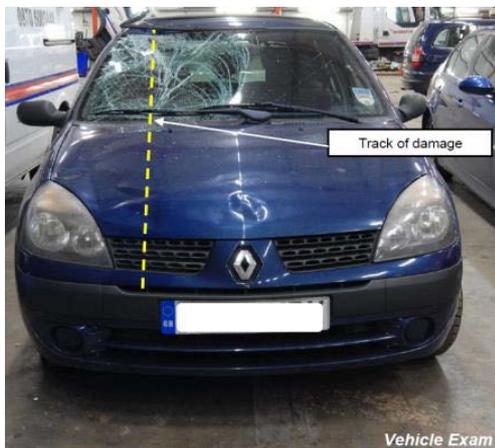 |

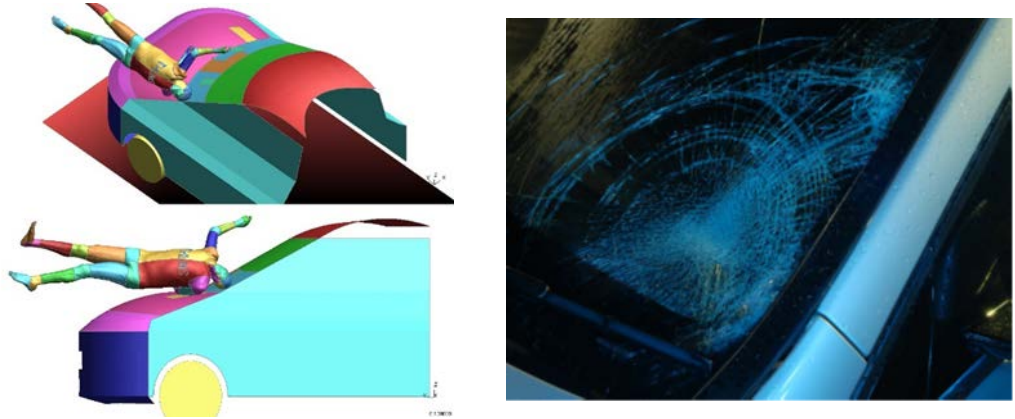

*Table 6: Vehicle damage and Pedestrian Kinematics [9][10]*

The vehicle geometries were reconstructed from blueprints and their respective local stiffness calibrated against EuroNCAP pedestrian test results [24][25][26].

4.2 Trauma Computation and Results

*Toyota Corolla Brain Trauma Results*

Step 1: Extraction of pedestrian kinematics and PVP during the accident

The accident was initially reconstructed according to the accident report, ensuring that the vehicle damage was consistent with the pedestrian kinematics. During this step, the PVP was extracted, as well as the white and grey matter velocities at the time of impact (Table 6). It could be noticed, in this instance, that these velocities at the moment of impact were different from the vehicle impact speed, as illustrated in Figure 8 and Figure 9.

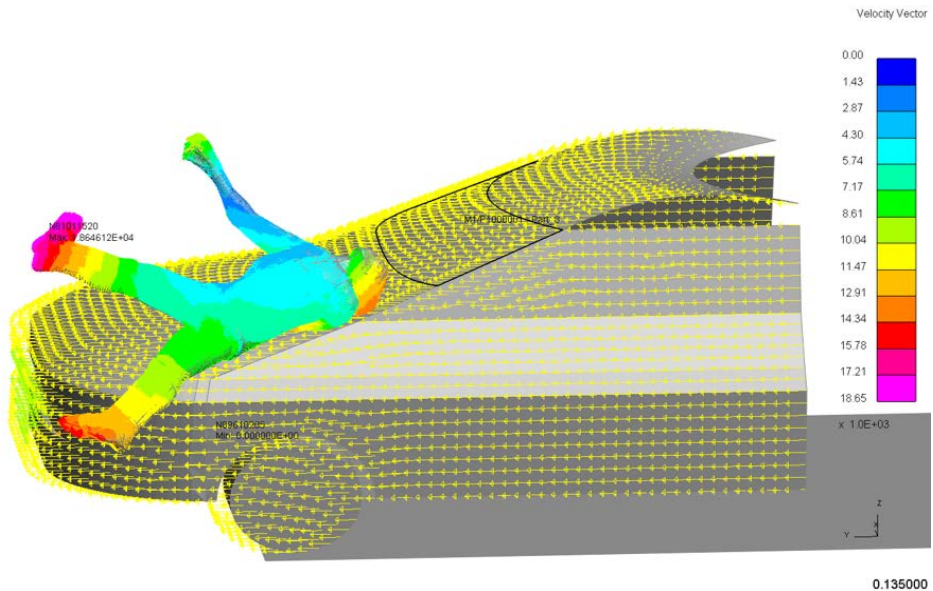

*Figure 8: Toyota Corolla - Collision Velocity Profile (mm/s)*

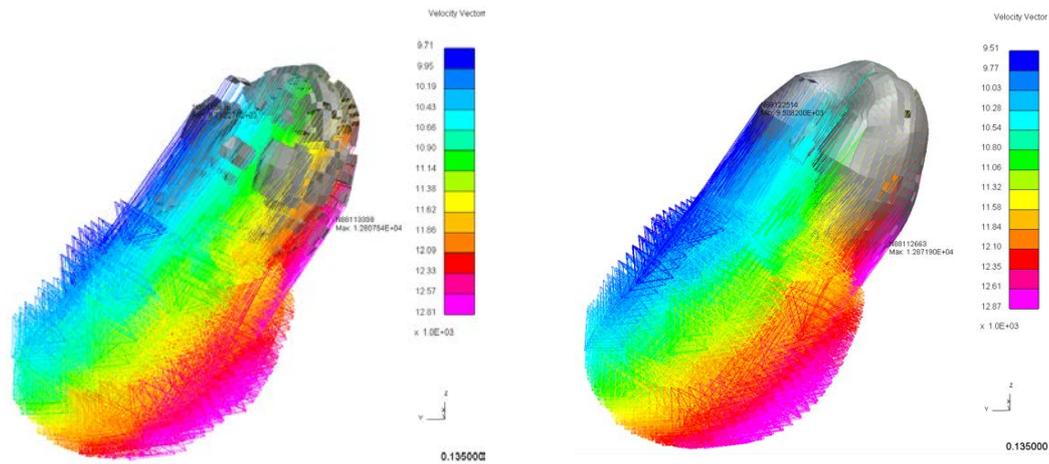

*Figure 9: Toyota Corolla - Brain velocity plot (Grey Matter (right), White Matter (left)) – Units (mm/s)*

| Organ | Resultant Velocity in car line (m/s) | Resultant velocity perpendicular to the windscreen (m/s) |
| --- | --- | --- |
| Grey Matter | 12.87 | 7.53 |
| White Matter | 12.81 | 7.89 |

*Table 7: Summary of Toyota Corolla brain velocities (at the time of impact)*

Step 2: Creation of the OTM model for this specific accident

The pedestrian kinematics was 'rewound' back in time, and repositioned 3mm from the bonnet surface, just prior to contact. This step is performed so that the pedestrian hits the same location of the vehicle (as the collision is unique). The pedestrian is then impacted at different speeds, respecting the direction vector of the pedestrian kinematics and impact location from Step 1, to construct an OTM model for each organ, comparable to Figure 4.

Step 3: Overlay step 1 and step 2 to extract trauma value

The initial AIS value, based on Equation 4, is computing injury severity based on a model with constant volume elements. Looking at Figure 10, it can be seen that for both grey and white matter, there is a potential of blood loss, as the MPS values are exceeding 25.5%. As a consequence, if volume decrease, AIS will increase. The AIS value computed using the PVP method will be therefore increased by '+1'.

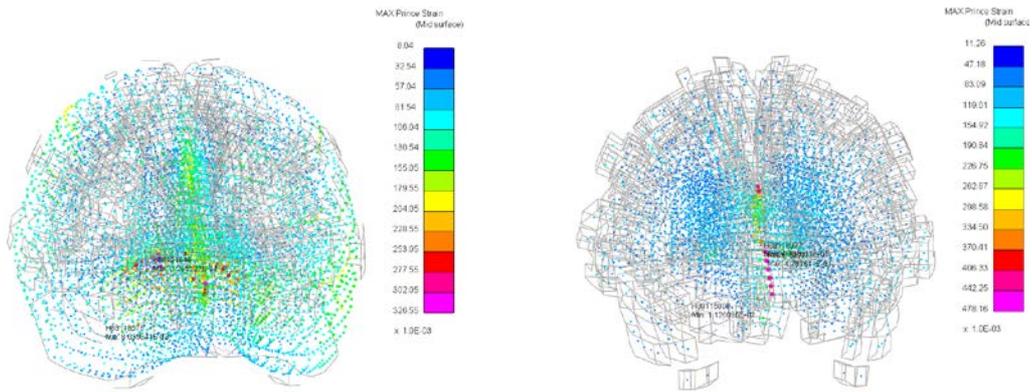

*Figure 10: Toyota. Maximum Principal Strain observed during the impact (Grey Matter – Left; White Matter – Right*

The white and grey matter brain velocities of the actual impact are remapped on the OTM trauma graphs, as shown in Figure 11 and Figure 12. For completeness, the results from the collision simulation including the full kinematics has also been included, in order to test whether the repositioning method was acceptable.

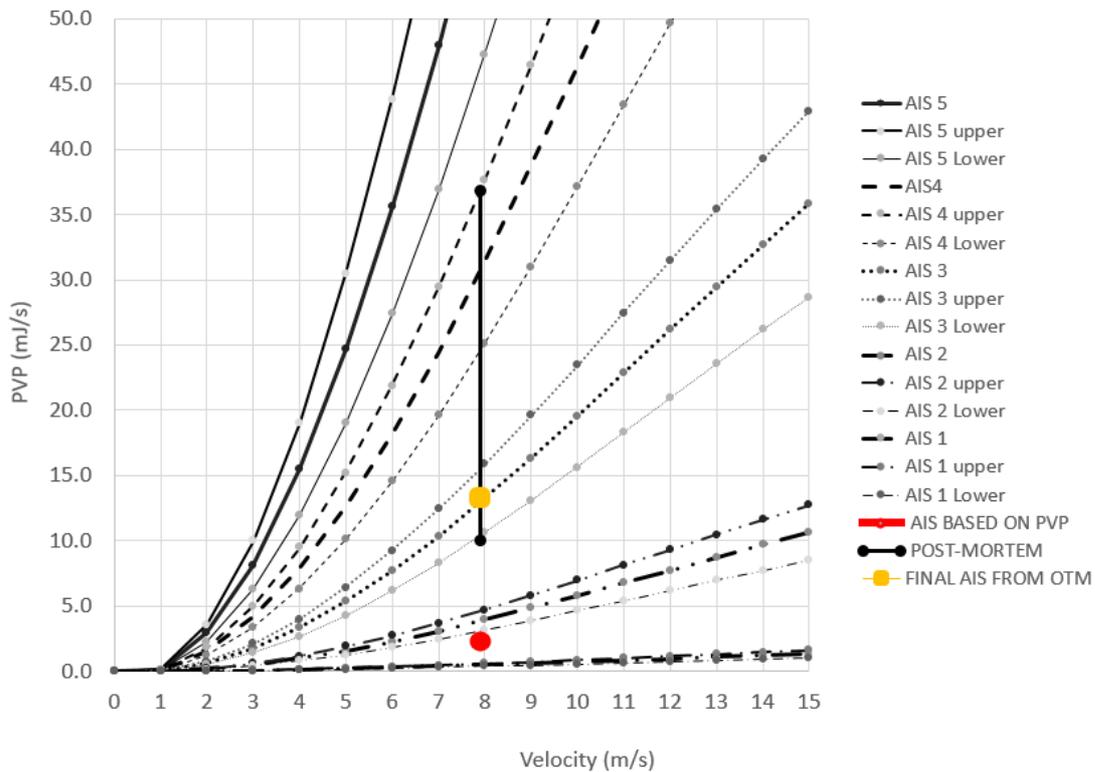

*Figure 11: Toyota Corolla - White Matter Trauma*

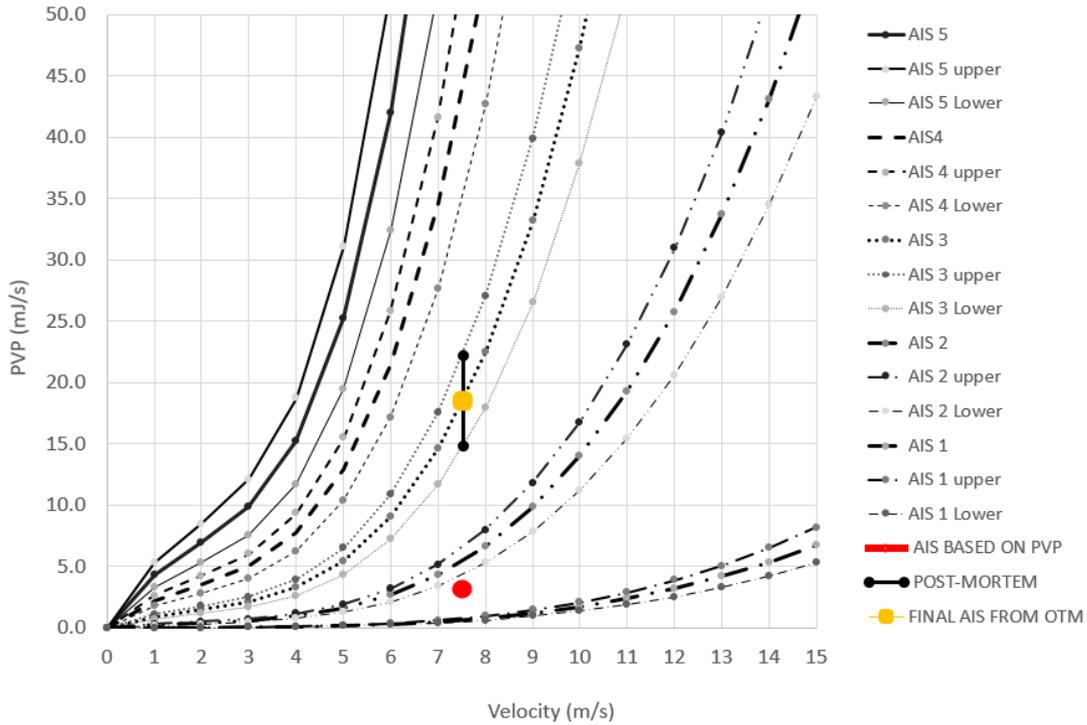

*Figure 12: Toyota Corolla - Grey Matter Trauma*

The collision impact speed was 11.2m/s, however the brain velocity was different at the time of impact. Consequently, the AIS values plotted (red dots) is adjusted to match the true organ speeds, Looking at Figure 11 and Figure 12, the AIS value for the white matter is 2 (at 7.89m/s) and the grey matter 1 (at 7.53m/s).

The process is repeated for Case 2 and Case 3. Their kinematics and trauma plots can be found in Appendix 3, 4, 5 and 6. The mathematical parameter fits for the three collisions are provided in Appendix 7.

In all the cases, the head injury predictions were similar to the Post Mortem results, as shown in Table 8. When no evidence was recorded in the PM, it did not necessarily mean that there is no injury, but that there is no observable injury. Consequently, no observation could mean that the AIS range could be from 0 to 2. This step has been taken, as it was found that, overall, the quality of autopsy reports (PM) is often questioned: just half of PM reports 52% (873/1,691) were considered satisfactory by experts, 19% (315/1,691) were good and only 4% (67/1,691) were excellent. Over a quarter were marked as poor or unacceptable. Proportionately, there were more reports rated 'unacceptable' for those cases that were performed in a local authority mortuary (21/214 for local authority mortuary cases versus 42/1,477 for hospital mortuary cases)" [27]. Consequently, for trauma injury severities cases not observed in the PM, a probable PM range has been included and is illustrated in Appendix 4 and 6. Appendix 8 is providing the trauma injury estimation using the standard THUMS recommended output, as per Table 2.

All the study results are listed in Table 8.

| Vehicle (Case id) | PM report details | Organs/Tissue | Injury | AIS from PM | CAE Prediction | MPS THUMS (Appendix 8) (AIS estimation) |
|---|---|---|---|---|---|---|
| Toyota Corolla (1) | Subarachnoid haemorrhage. The brain appeared diffusely swollen to a mild degree. There were contusions on the inferior aspect of the right temporal lobe. | White Matter | Diffuse Axon Injury (just reached) | 3 - 4 | 3 (2+1) | 48% (AIS 4 > 21%) |
| | | Grey Matter | Brain Contusion | 3 | 3 (2+1) | 32% (AIS 3 > 26%) |
| Renault Clio (2) | No evidence of skull fracture and brain showed no evidence of contusion | White Matter | No evidence | 0-2 | 2 (1+1) | 127% (AIS 4 > 21%) |
| | | Grey Matter | No contusion | 0-2 | 2 (1+1) | 113% (AIS 3 > 26%) |
| Mercedes Benz (3) | Multiple area of cerebral contusion Rupture at right parietal lobe Cerebral oedema Subarachnoid haemorrhage Subdural haemorrhage | White Matter | Diffuse Axon Injury | 3 – 4 | 3 (2+1) | 72% (AIS 4 > 21%) |
| | | Grey Matter | Brain Contusion | 3 | 2 (1+1) | 58% (AIS 3 > 26%) |

*Table 8: Study results for brain injuries*

## *5. Discussion*

In Case 1 (Table 8), the PM is stating that subarachnoid haemorrhage was observed in the white matter and the brain appeared diffusely swollen to a mild degree", which suggests that the DAI has just been reached, hence the white matter PM AIS has to be at least a 3. The MPS method is suggesting at least an AIS 4. The grey matter MPS predictions were accurate (AIS 3). For the Toyota, the injury severity for the white and grey matter were both also computed as AIS 3. This severity was calculated by adding

'1' AIS to the AIS 2 initially computed by the OTM because SDH was evidenced, i.e. MPS > 25.5%, in Appendix 8. This is a good match to the PM.

Considering Table 8, in the case of the Renault Clio, the OTM model suggests a minor injury (AIS 1), which is compatible with the PM, however the MPS levels suggest that SDH occurred (Appendix 8), hence an injury severity of AIS 2. The PM did not record any blood loss, indicating that the PM may be questionable.

Finally, in the case of the Mercedes Benz the white matter and grey matter the threat to life were computed to be AIS 3 and AIS 2 respectively. The initial injuries severities were calculated as AIS 2 and AIS 1, and were then increased by +1 because haematoma was observed on both white and grey matter.

It can be observed that the comments from the PM, which are the interpretations from the forensic pathologist, and that the AIS level is in this instance a function of how much blood is observed during the autopsy. The THUMS model is using a Lagrangian method, which implies that the volume of each element remains constant during the impact. This method cannot cater for bleeding. Including bleeding would involve a reformulation of the THUMS' brain model and include Smooth Particle Hydrodynamics (SPH) or Arbitrary Lagrangian and Eulerian (ALE) formulations. Consequently, the AIS under-prediction using PVP is a logical numerical outcome in the case of blood loss.

Looking at all these results, it can be observed that the MPS method does not allow the grading of AIS as a function of MPS level. Only one level is provided, i.e. the critical one, which is a serious limitation when trying to match PM to computations. The MPS overall over-estimates the injury, while PVP under-predicts should bleeding occur. This study is suggesting that maybe a new brain model would be necessary to capture the bleeding effect which is recorded in the PMs.

These results also may be sensitive to the geometry of the vehicle model. Indeed, the vehicle model shape was extracted from blueprints. In the future, it would be maybe necessary to obtain a scanned surface of the vehicle so that the exact curvature and the local geometry are accurately captured. Also, the vehicle stiffness was based on calibrating the head impact zone using a head impactor HIC panel thickness calibration to match the local pedestrian EuroNCAP performance rating [24][25][26]. Maybe another method of vehicle modelling, for example using the APROSYS bonnet stiffness corridors, would be another venue of investigation.

An important parameter, is that it is not known whether each of these accidents involved a head impact to the ground, which would increase the head AIS level. In all cases, the trauma caused by the primary impact is always the same or lower than the trauma at the end of the collisions. Consequently, if the PVP method is under-predicting in the primary impact, the trauma severity outcome discrepancy could have come from the pedestrian's head landing on the ground.

## *6. Conclusions*

The research has produced a unique method to compute the different levels of trauma severity in the brain white and grey matter. Unlike the standard Maximum Principal Strain (MPS) method, which can only state whether a critical injury severity has been reached, this new Organ Trauma Model (OTM) provides the capabilities to extract the full range of AIS levels (1 to 5), which is a unique feature.

The work has also implemented ageing in a THUMS human body model (HBM), by altering the mechanical properties of its soft and hard tissues, scaling their values to include body degeneration as a function of ageing. This aged model, which can be customised to represent human from the age of 20 to 80 years old, is also a unique feature of this research. When the THUMS aged model was tested against a rigid pendulum using the OTM method, it was observed that the impact capability of older people is less than younger ones: this statement is compatible with what is observed in real life.

During this work and deriving the OTM fundamental mathematical equations, it became apparent that a Lagrangian formulation of any Finite Element based HBM has no capability to model accurately bleeding, as the elements have a constant volume during the whole duration of the computation. To overcome this limitation the OTM method has included a correction factor in this study to consider subdural Hematoma (SDH), based on 25.5% MPS threshold [7]. It has been hypothesised that the final AIS value to be the addition of '+1' AIS level to the AIS obtained by the OTM method. This AIS value increase choice was made as it was plausible based on the limited number of accident data. In the future, more cases are necessary to statistically refine this hypothesis.

In any case, when comparing the two methods against three accidents, it was observed that the OTM PM's predictions were more accurate than MPS. The MPS method predicted AIS 5, when the Post Mortem (PM) suggested no injuries. Overall the PMs computed by the OTM method are plausible, suggesting that the OTM opens up a new way of assessing human brain injury severity and provide additional granularity above and beyond present methods.

The research has highlighted that there may be a need to review the brain model and include means to model bleeding, maybe by adding a Smooth Particle Hydrodynamics (SPH) or Arbitrary Lagrangian and Eulerian (ALE) formulation. It is believed that the OTM method can be also used for thorax and abdomen soft tissue organs.

## *7. Further Work*

It is proposed to revisit the vehicle geometry and stiffness characteristics to refine the accuracy of the OTM model. Following this, the work will be extended to analyse the thorax and abdomen organ injury severity responses using the same method, as well as contact more partners to increase the data samples to further test the OTM method.

More investigation is needed to model bleeding in the THUMS model.

## *Acknowledgements*

The authors would like to thank the Road Safety Trust [27] for their unconditional support and for funding this research Road Safety Trust (RST 65-3-2017) "Reducing Road Traffic Casualties through Improved Forensic Techniques and Vehicle Design ("RoaD"), as well as the UK Police Force and the UK Coroners who made this research possible.

# *Appendix 1: Algebraic Derivation of Trauma Severity (1/5)*

This section is a mathematical proof to the derivation of the trauma severity, based on Peak Virtual Power. Three scenarios will considered: (1) pedestrian is deformable and the vehicle rigid; (2) vehicle and pedestrian sharing the same criteria; and (3) vehicle and pedestrian sharing different criteria in order to derive trauma severity. The rigour and validity of the final derivation of equation (3) will be verified by confirming that (1) and (2) can be re-derived by reducing the impact assumptions.

The proof concludes that Trauma is fundamentally proportional to the square of the velocity (also proven by accident data, $R^2 > 0.9$). The use of an existing empirical data set for pedestrian accidents has shown that this relationship maybe further refined by including an additional velocity term to provide a cubic relationship ($R^2 > 0.95$). The authors chose the cubic approach for the work reported in this paper, while the squared relationship would have also been acceptable.

1. Assuming that the pedestrian is deformable and the vehicle rigid

The impact kinetic energy of the pedestrian at $t_0$, is converted into strain energy (deformation) and kinetic energy. This is a time dependant relationship. $V_{t0}$ is the impact speed, while $\sigma_t$ and $v_t$ are the stress and speed generated inside the system as the time passes.

$$\frac{1}{2}mv_{t_0}^2 = \frac{\sigma_t^2}{2E}vol + \frac{1}{2}mv_t^2$$

VP (virtual power) is the product of the stress and the strain rate. If the equation above is rearranged to make stress the subject and then multiplied through by strain rate the result is an equation for VP that is time dependent:

$$\sigma\dot{\varepsilon} = VP(t) = \frac{v_t}{L}\sqrt{\rho E\left(v_{t_0}^2 - v_t^2\right)}$$

PVP is the maximum value of VP(t). VP(t) is maximum (proven in Appendix 2) when:

$$v_{t_0} = v_t\sqrt{2}$$

Giving

$$PVP = \frac{v_{t_0}}{\sqrt{2}L}\sqrt{\rho E\left(v_{t_0}^2 - \left[\frac{v_{t_0}}{\sqrt{2}}\right]^2\right)}$$

Reducing to

$$PVP = \frac{1}{2L}\sqrt{\rho E}\,v_{t_0}^2$$

<u>Conclusion:</u> In this configuration, trauma severity, or PVP, is proportional to the square of the velocity (aligned with impact direction), however the trauma severity would be lower if the vehicle is not rigid, i.e. deforms, hence the next formulation, considering the vehicle deformable.

## Appendix 1: Algebraic Derivation of Trauma Severity (2/5)

2. Assuming that the pedestrian and the vehicle share the same stiffness characteristics

If a constant stiffness 'E' is assumed for the vehicle and the pedestrian, then:

$$\frac{1}{2}mv_{t_0}^2 = \frac{\sigma_t^2}{2E}vol + \frac{\sigma_t^2}{2E}vol + \frac{1}{2}mv_t^2$$

This then becomes

$$\sigma = \sqrt{\frac{m(v_{t_0}^2 - v_t^2)}{2*\frac{vol}{E}}}$$

Taking account that density = mass / volume, then:

$$\sigma = \sqrt{\frac{m(v_{t_0}^2 - v_t^2)}{2*\frac{m}{E\rho}}}$$

Reducing to:

$$\sigma = \sqrt{\frac{\rho E(v_{t_0}^2 - v_t^2)}{2}}$$

Hence, if the collision partners share the same characteristics the stress, in comparison to the previous example, is reduced by $1/\sqrt{2}$

As VP is the product of the stress and the strain rate:

$$\sigma\dot{\varepsilon} = VP = \frac{v_t}{L}\sqrt{\frac{\rho E(v_{t_0}^2 - v_t^2)}{2}}$$

Note that the strain rate is the same for both collision partners in this example. Further the maximum value can be found (Appendix 2) when:

$$v_{t_0} = v_t\sqrt{2}$$

Giving

$$PVP = \frac{1}{2L}\sqrt{\frac{\rho E}{2}}v_{t_0}^2$$

Again, trauma severity is proportional to the square of the impact velocity (aligned with impact direction). It can be here noted that when two partners that both deform, that PVP is lower, which is logical. The stiffness of the vehicle is therefore important.

## *Appendix 1: Algebraic Derivation of Trauma Severity (3/5)*

3. Assuming that the pedestrian and the vehicle share the different stiffness characteristics

Assuming that the collision partners have different stiffness characteristics (c is car and p is pedestrian):

$$\frac{1}{2}m{v_{t_0}}^2 = \frac{\sigma_t^2}{2E_c}vol_c + \frac{\sigma_t^2}{2E_p}vol_p + \frac{1}{2}m{v_t}^2$$

The full equation where both collision partners are deformable and have different values of E is shown below:

$$\sigma = \sqrt{\frac{m\left({v_{t_0}}^2 - {v_t}^2\right)}{\frac{vol_c}{E_c} + \frac{vol_p}{E_p}}}$$

VP is the product of the stress and the strain rate then for the pedestrian, hence:

$$\sigma\dot{\varepsilon}_p = \text{VP} = \frac{v_t}{L_p}\sqrt{\frac{m_p\left({v_{t_0}}^2 - {v_t}^2\right)}{\frac{vol_c}{E_c} + \frac{vol_p}{E_p}}}$$

Taking account that density = mass / volume.

$$\text{VP} = \frac{v_t}{L_p}\sqrt{\frac{m_p\left({v_{t_0}}^2 - {v_t}^2\right)}{\frac{m_c}{\rho_c E_c} + \frac{m_p}{\rho_p E_p}}}$$

VP tends to a maximum value (appendix 2) when:

$$v_{t_0} = v_t\sqrt{2}$$

Giving:

$$PVP = \frac{1}{2L_p}\sqrt{\left(\frac{\rho_p E_p \rho_c E_c m_p}{\rho_p E_p m_c + \rho_c E_c m_p}\right){v_{t_0}}^2}$$

Again, trauma severity is proportional to the square of the vehicle impact velocity (aligned with the impact direction).

## *Appendix 1: Algebraic Derivation of Trauma Severity (4/5)*

Assuming that L is the ratio between the contact area A and the volume V of the pedestrian.

$$AIS \propto PVP = \frac{A_p}{2V_p}\sqrt{\left(\frac{\rho_p E_p \rho_c E_c m_p}{\rho_p E_p m_c + \rho_c E_c m_p}\right)v_{t_0}^2}$$

The equation above is the generic algebraic formulation of trauma severity.

It can be noted that the injury severity is also dependant on the contact area, hence the vehicle profile.

<u>Conclusion AIS is function of:</u>
- the contact area between the vehicle and the pedestrian
- Volume of material supporting the impacted Area impacted
- pedestrian mass, density and stiffness
- vehicle mass, density and stiffness
- Impact speed (speed orthogonal to the impacted structure)
- Ageing, as material properties and volume are age dependant.

4. Verification of the generic algebraic formulation of trauma severity

*If both partners have the same characteristics (section 1), then the generic equation reduces to:*

$$PVP = \frac{1}{2L_p}\sqrt{\frac{\rho E}{2}v_{t_0}^2}$$

*If we refer back to previous example (section 2) and assume that collision partner is rigid then*

$$PVP = \frac{1}{2L_p}\sqrt{\left(\frac{\rho_p E_p \rho_c m_p}{\frac{\rho_p E_p m_c}{E_c} + \rho_c m_p}\right)v_{t_0}^2}$$

As $E_c$ tends to infinity, the term containing $E_c$ tends to zero and can be discarded, hence:

$$PVP = \frac{1}{2L_p}\sqrt{\left(\frac{\rho_p E_p \rho_c m_p}{\rho_c m_p}\right)v_{t_0}^2}$$

Or

$$PVP = \frac{1}{2L_p}\sqrt{\rho_p E_p}\, v_{t_0}^2$$

# Appendix 1: Algebraic Derivation of Trauma Severity (5/5)

5. Generic algebraic formulation of trauma severity and real-life accident evidence

The generic equation below is the Generic algebraic formulation of trauma severity:

$$AIS \propto PVP = \frac{A_p}{2V_p}\sqrt{\left(\frac{\rho_p E_p \rho_c E_c m_p}{\rho_p E_p m_c + \rho_c E_c m_p}\right)}\, v_{t_0}^{\,2}$$

This equation is compatible with real-life accident published evidence illustrated in Figure 13.

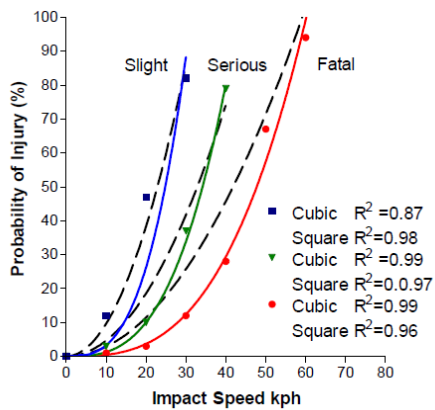

Figure 13: Pedestrian accident cases: relationship between threat to life (AIS) and vehicle impact speed [5].

It can noticed that for serious (AIS3) and fatal accident (AIS4+) that the correlation exponent is at least 0.96 for a squared interpolation and 0.99 for a cubic interpolation. This is showing that there is already a very strong relationship between AIS and the square of the impact velocity, hence proving the generic algebraic formulation of trauma severity derivation is reasonable and representative of the trauma severity representation phenomenon. In the case of fatal injuries, the polynomial fit is already very good (0.96), however it is even more accurate with a cubic exponent, suggesting that the cubic formation can capture more accurately, but marginally more accurately than a squared relationship, the real-life events of the fatality phenomenon, due to the simplification of the equations, assuming a linearity of stiffness and disregarding the non-linear visco-elastic material responses. Looking at Appendix 7, the cubic term is very low, hence this suggest a refinement in the prediction, but not of a fundamental relevance. Hence as the calibration will be based on AIS4, the authors have decided to use a cubic interpolation. Note that this choice does not void the validity of the generic algebraic formulation of trauma severity formulation derived in Appendix 1.

$$AIS \propto PVP = \frac{A_p}{2V_p}\sqrt{\left(\frac{\rho_p E_p \rho_c E_c m_p}{\rho_p E_p m_c + \rho_c E_c m_p}\right)}\, v_{t_0}^{\,3}$$

## *Appendix 2: Derivation of the maximum of VP.*

Finding the maximum of:

$$\sigma \dot{\varepsilon} = \text{VP} = \frac{v_t}{L}\sqrt{\frac{\rho E\left(v_{t_0}^2 - v_t^2\right)}{2}}$$

This equation can be re-written as:

$$\text{VP} \propto v_t^2\left(v_{t_0}^2 - v_t^2\right)$$

Therefore

$$\text{VP} \propto -v_t^4 + v_t^2 \cdot v_{t_0}^2$$

The maximum can be found by differentiating against $v_t^2$

$$\frac{d\left(v_t^4 - v_t^2 \cdot v_{t_0}^2\right)}{dv_t^2} = 0$$

Giving:

$$2 \cdot v_t^2 = v_{t_0}^2$$

Hence:

$$\boxed{v_{t_0} = v_t\sqrt{2}}$$

## Appendix 3: Velocity Plots – Case 2 (Renault Clio)

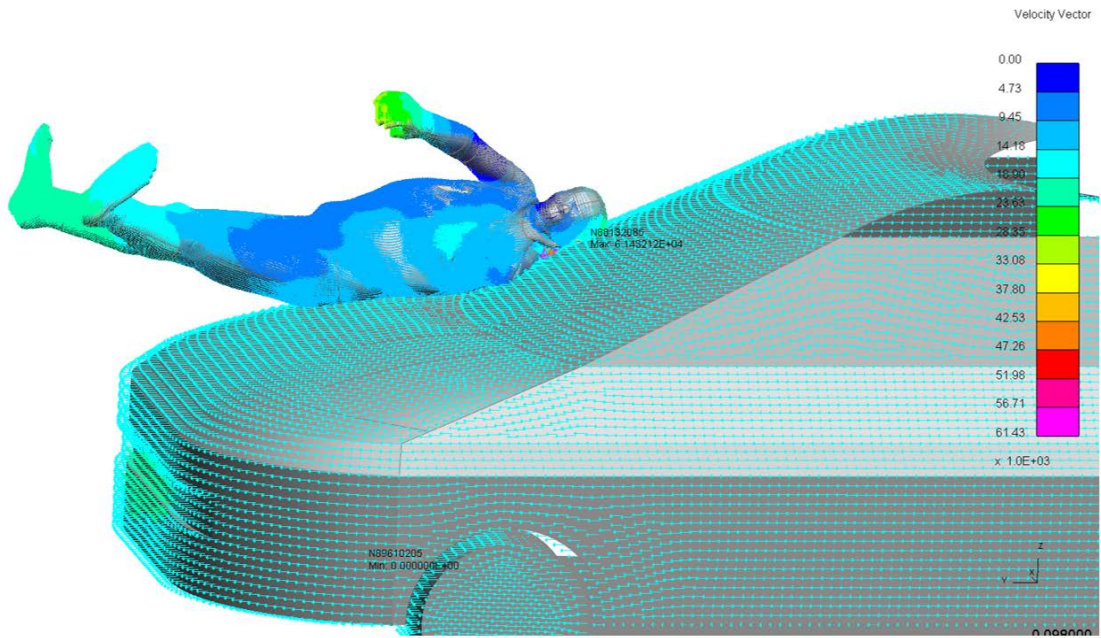

*Figure 14: Renault Clio - Collision Velocity Pattern (mm/s)*

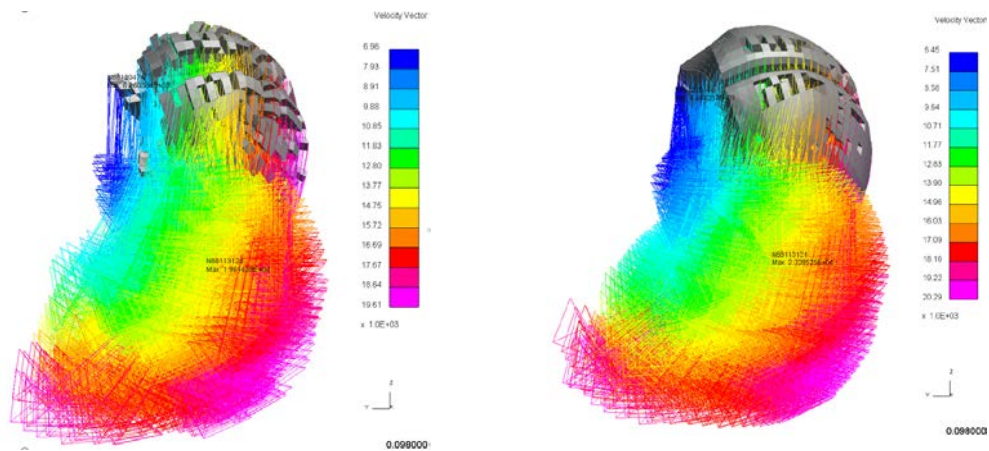

*Figure 15: Renault Clio - Brain velocity plot (White Matter (left), Grey Matter (right)*

| Organ | Resultant Velocity in car line (m/s) | Resultant velocity perpendicular to the windscreen (m/s) | Time (s) |
| --- | --- | --- | --- |
| Grey Matter | 20.29 | 17.35 | 0.0980 |
| White Matter | 19.61 | 16.15 | 0.0980 |

*Table 9: Summary of Renault Clio brain velocities*

# Appendix 4: Trauma Plots – Case 2 (Renault Clio)

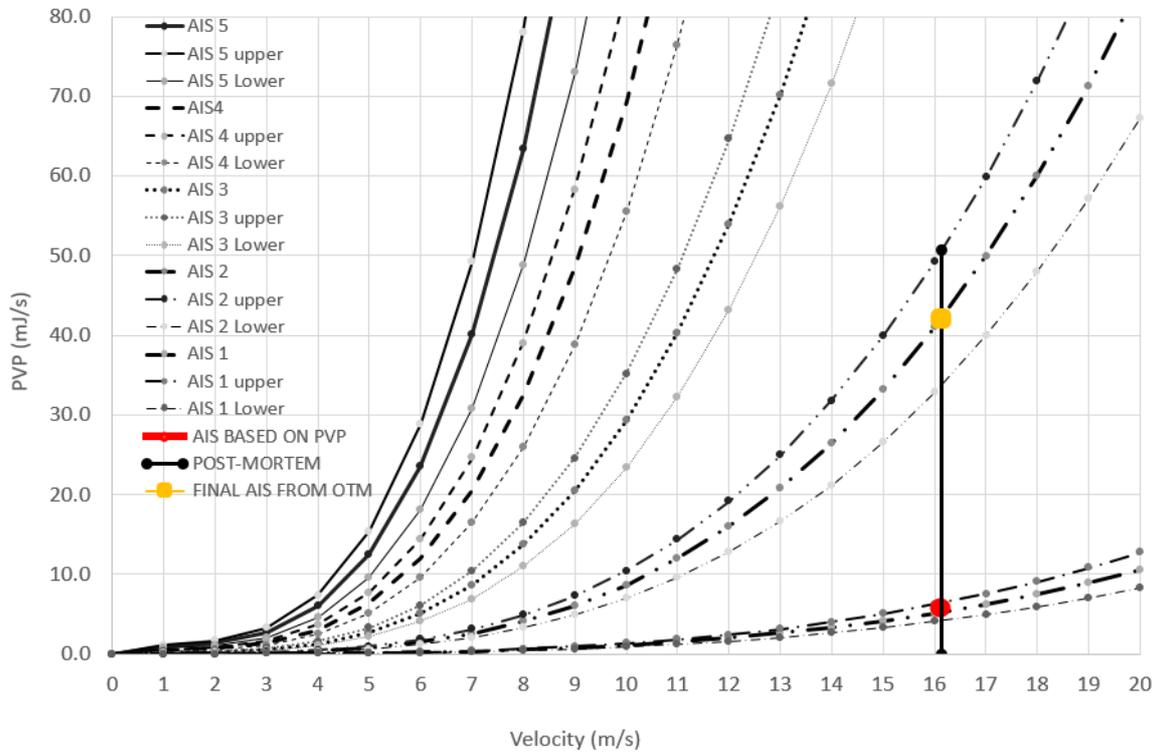

*Figure 16: Renault Clio - White Matter Trauma*

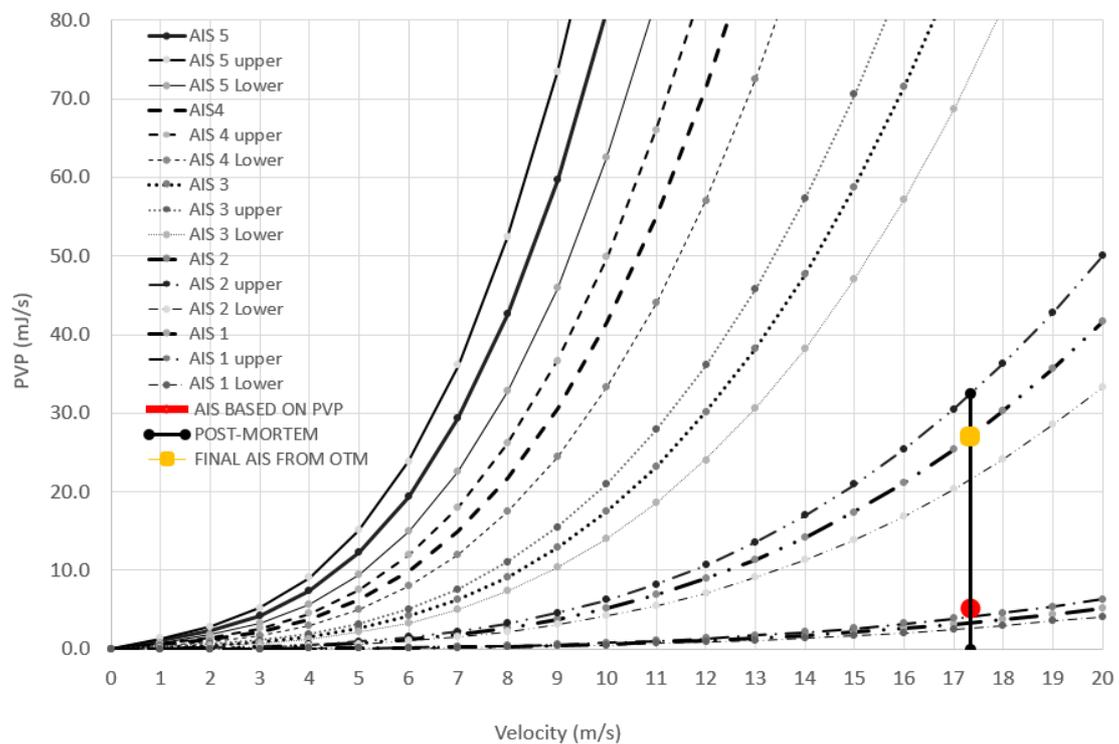

*Figure 17: Renault Clio - Grey Matter Trauma*

# Appendix 5: Velocity Plots – Case 3 (Mercedes Benz)

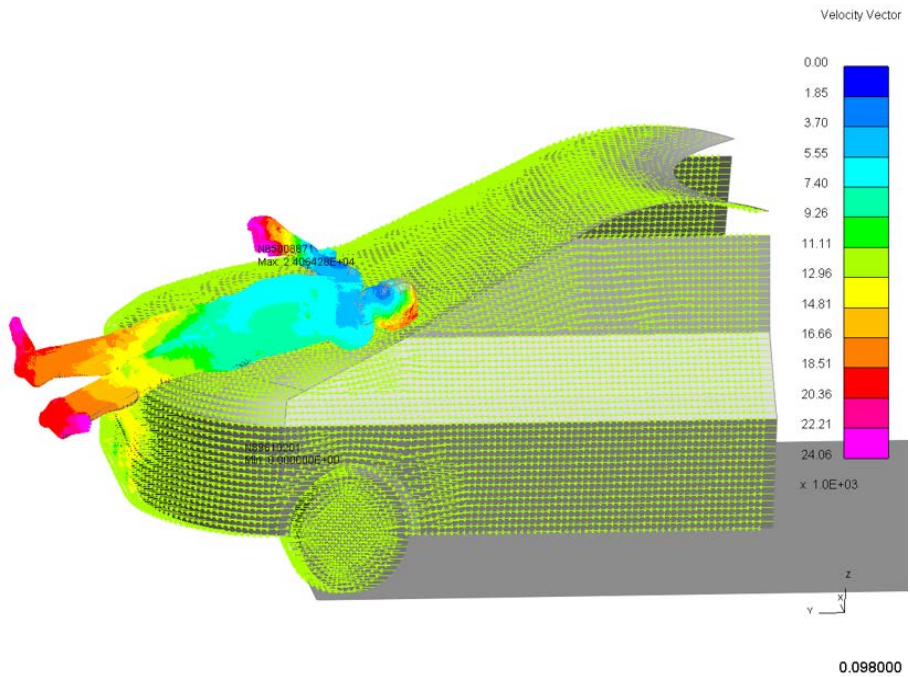

*Figure 18: Mercedes Benz - Collision Velocity Pattern (mm/s)*

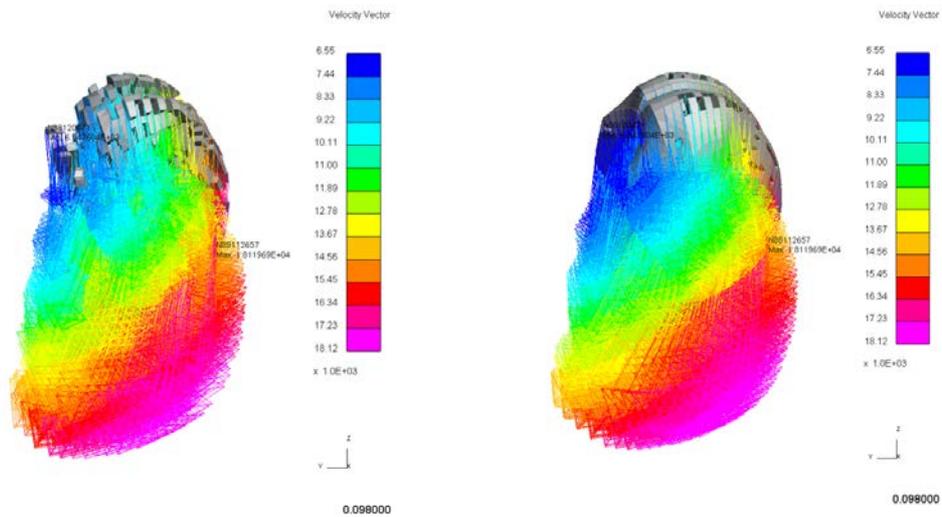

*Figure 19: Mercedes Benz - Brain velocity plot (White Matter (left), Grey Matter (right)*

| Organ | Resultant Velocity in car line (m/s) | Resultant velocity perpendicular to the windscreen (m/s) | Time (s) |
|---|---|---|---|
| Grey Matter | 18.12 | 17.34 | 0.0980 |
| White Matter | 18.12 | 16.15 | 0.0980 |

*Table 10: Summary of Renault Clio brain velocities*

## Appendix 6: Trauma Plots – Case 3 (Mercedes Benz)

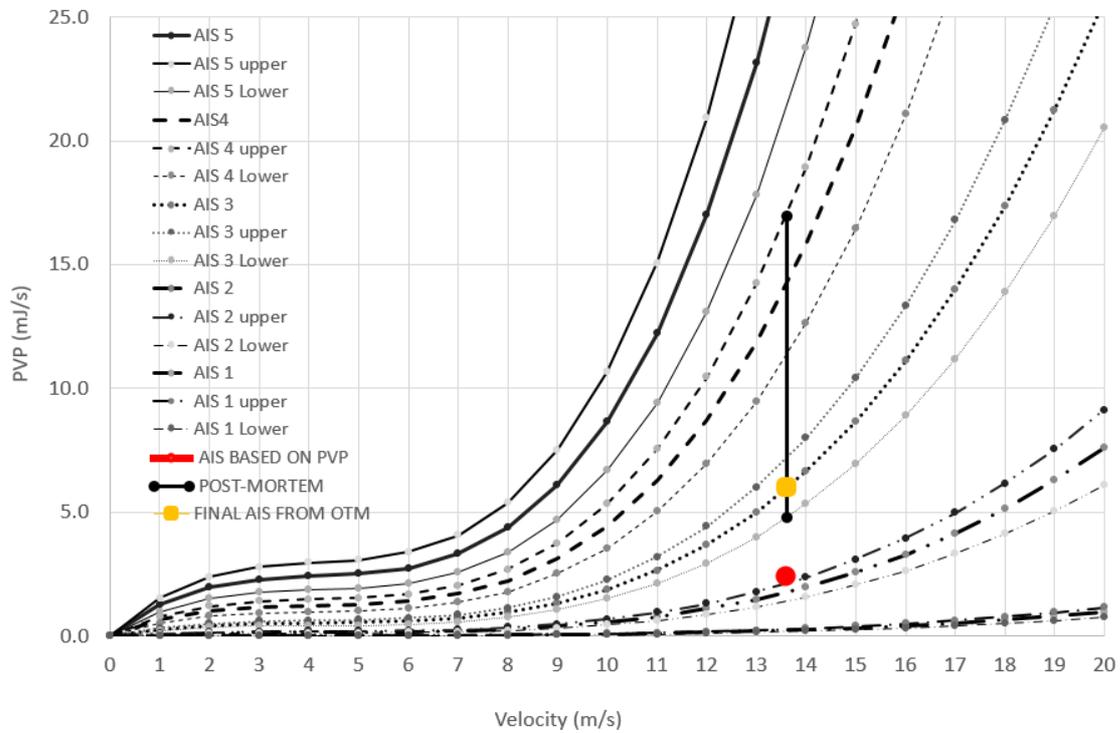

*Figure 20: Mercedes Benz - White Matter Trauma*

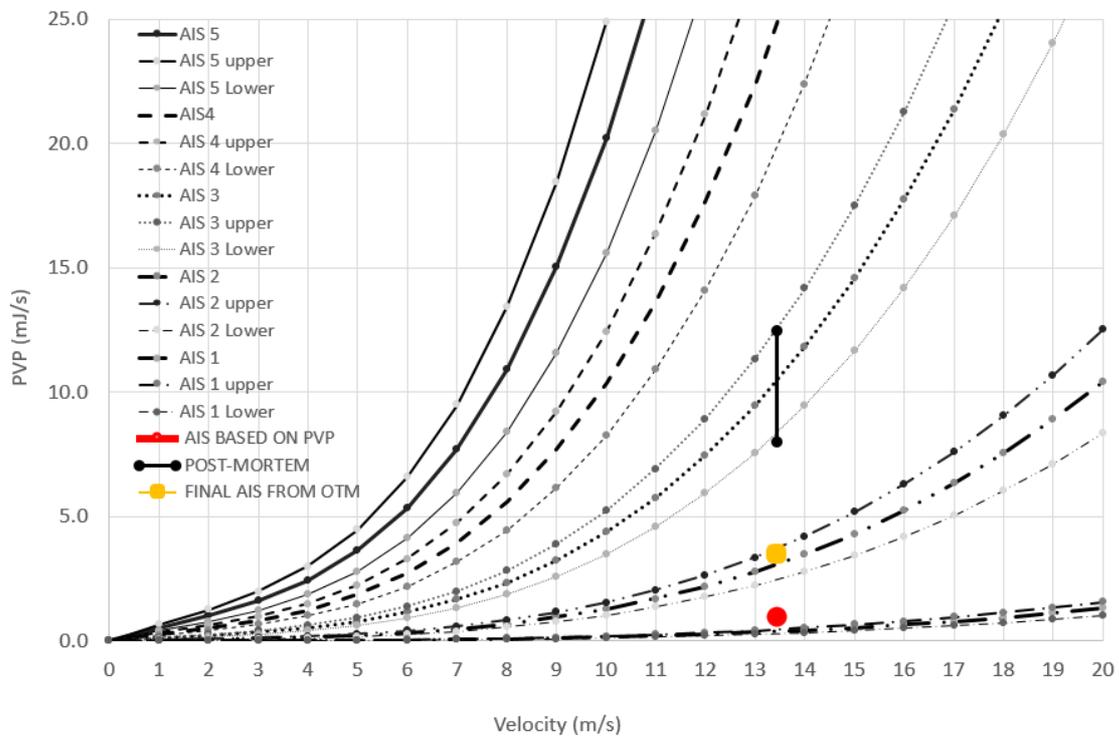

*Figure 21: Mercedes Benz - Grey Matter Trauma*

## Appendix 7: Mathematical Fits (AIS4) for the 3 Collisions

| Trauma Calibration Parameter Values | | | | |
|---|---|---|---|---|
| Parts Identifier (White Matter) – right hand side | white_matter_cerebrum_r | 88000100 | | |
| Parts Identifier (White Matter) – left hand side | white_matter_cerebrum_l | 88000120 | | |
| Parts Identifier (Grey Matter) – right hand side | gray_matter_cerebrum_r | 88000101 | | |
| Parts Identifier (Grey Matter) – left hand side | gray_matter_cerebrum_l | 88000121 | | |
| PVP = $a \cdot V^3 + b \cdot V^2 + c \cdot V$ | | | | |
| Parameter Values | | a | b | c |
| Case 1: Toyota Corolla | White matter | -0.0217 | 0.746 | -0.6537 |
| | Grey matter | 0.0765 | -0.4207 | 1.2828 |
| Case 2: Renault Clio | White matter | 0.1025 | -0.4064 | 0.7509 |
| | Grey matter | 0.0765 | -0.4207 | 1.2828 |
| Case 3: Mercedes Benz | White matter | 0.0148 | -0.1844 | 0.8078 |
| | Grey matter | 0.0051 | -0.0206 | 0.133 |

# Appendix 8: Maximum Principal Strain Responses

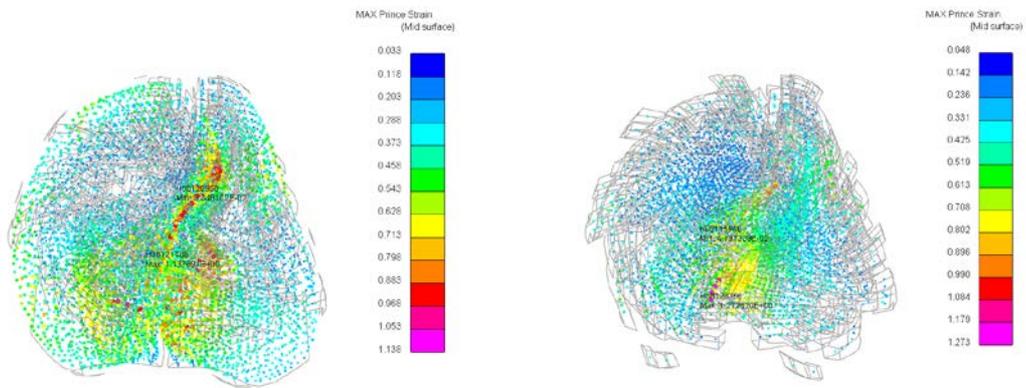

*Figure 22: Renault Clio. Maximum Principal Strain observed during the impact (Grey Matter – Left; White Matter – Right*

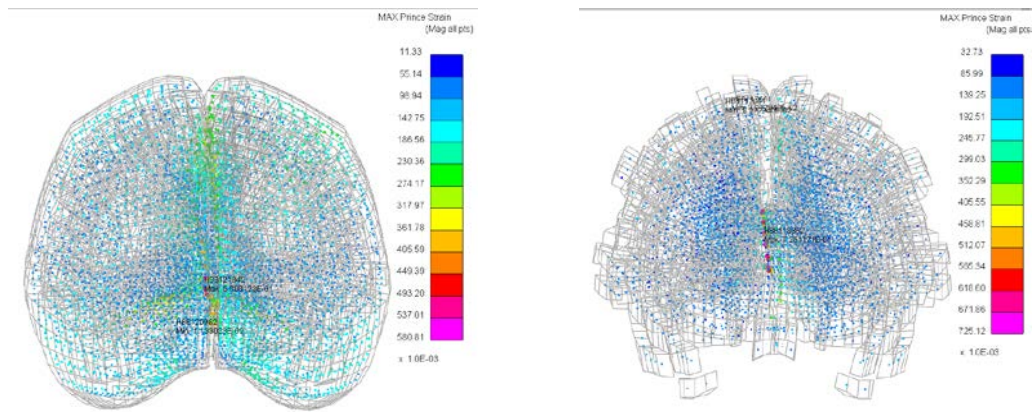

*Figure 23: Mercedes Benz. Maximum Principal Strain observed during the impact (Grey Matter – Left; White Matter – Right)*